\documentclass[pra,reprint,showpacs,showkeys,showpacs,times,superscriptaddress]{revtex4-1}
\usepackage{amsmath,amssymb,amsthm,times,graphics,graphicx,bm,subfigure}

\usepackage[colorlinks,linkcolor=red,citecolor=blue,urlcolor=blue]{hyperref}

\def\AD{\textcolor{black}}
\def\PC{\textcolor{blue}}

\newcommand{\be}{\begin{equation}}
\newcommand{\ee}{\end{equation}}
\newcommand{\ben}{\begin{eqnarray}}
\newcommand{\een}{\end{eqnarray}}
\newcommand{\bes}{\begin{subequations}}
\newcommand{\ees}{\end{subequations}}
\newcommand{\bF}{\begin{figure}}
\newcommand{\eF}{\end{figure}}

\newcommand{\arxiv}[2][arxiv:]{\href{http://arxiv.org/abs/#1#2}{#1#2}}

\def\ket#1{ | #1 \rangle}
\def\bra#1{{\langle #1 |  }}
\def\vec#1{\boldsymbol{#1}}
\newcommand{\avg}[1]{\langle #1 \rangle}
\def\braket#1#2{\langle #1 |  #2 \rangle}
\def\tr#1{{\rm{Tr}}\left[#1\right]}

\def\pd2v#1#2#3{\frac{\partial^2 #1}{\partial #2 \partial #3}}

\def\absq#1{\left| #1 \right|^2}

\def\binom#1#2{\left( \begin{array}{c} #1 \\ #2 \end{array} \right)}

\def \cv{\mathrm{Cov}}

\def \I#1{\mathcal{I}_{ #1}}

\def \vec#1{\mathbf{#1}}
\def \2x2mat#1#2#3#4{
\left( \begin{array}{cc}
#1 &  #2 \\  #3 &  #4
\end{array} \right)
}
\newcommand{\proj}[1]{\mbox{$|#1\rangle \!\langle #1 |$}}

\def \i{i}

\def \IF{\mathcal{I}_{\mu\nu}\left[\rho\right]}
\def\pket{ | \psi_l \rangle}
\def\dpket{ | \partial_\mu \psi_l \rangle}
\def\pbra{{\langle \psi_l |  }}
\def\dpbra{{\langle \partial_\mu \psi_l |  }}

\def \dlpl{ \partial_\mu \log p_l }
\def \dpr { \partial_\mu \rho }
\def \ri {\rho_l}

\def \lmi {L_{l,\mu}}

\begin{document}

\title{A tradeoff in simultaneous quantum-limited phase and loss estimation in interferometry}
\author{Philip J. D. Crowley}
\address{London Centre for Nanotechnology, University College London, Gordon St., London, WC1H 0AH, United Kingdom}
\author{Animesh Datta, Marco Barbieri and I.A. Walmsley}
\address{Clarendon Laboratory, Department of Physics, University of Oxford, OX1 3PU, United Kingdom}


\begin{abstract}
Interferometry with quantum light is known to provide enhanced precision for estimating a single phase.
\AD{However, depending on the parameters involved, the quantum limit for the simultaneous estimation of multiple parameters may not attainable, leading to trade-offs in the attainable precisions.} Here we study the simultaneous estimation of two parameters related to optical interferometry: phase and loss, using a fixed number of photons. We derive a trade-off in the estimation of these two parameters which shows that, in contrast to single-parameter estimation, it is impossible to design a strategy saturating the quantum Cram\'{e}r-Rao bound for loss and phase estimation in a single setup simultaneously. We design optimal quantum states with a fixed number of photons achieving the best possible simultaneous precisions. Our results reveal general features about concurrently estimating Hamiltonian and dissipative parameters, and has implications for sophisticated sensing scenarios such as quantum imaging.
\end{abstract}

\maketitle

\section{Introduction}

High accuracy measurements of optical path differences, characterized by a phase difference $\phi$ between the two modes of the apparatus, have found use in many fields~\cite{Steel_1983,Ye_SH03,Abbott_2005,Yang_WHBDF01}, and has consistently featured as an important tool in physics, from the null result in the search for the drag of luminiferous aether~\cite{Michelson_1887} to detecting small variations in the refractive index of biological solutions~\cite{Luff98,Watts10,Pierangelo:11,Crespi11}. Most of these experiments are concerned with the estimation of a single parameter, the phase difference~\cite{Holland_1993,Braunstein_1994,Bollinger_1996,Lee_Kok_Dowling_2002,Dorner_et_al_2009,Datta_2011}. Indeed, a vast majority of metrological problems, particularly those associated with Hamiltonian dynamics, can be recast into one of phase estimation~\cite{Giovannetti_2004,Giovannetti_Lloyd_Maccone_2011}. For this task, quantum states of entangled photons are subject to a more favourable limit on precision compared to the best possible classical strategies~\cite{Mitchell_2003,Thomas-Peter_2011}. This quantum limit for phase estimation can always be attained~\cite{Braunstein_1994}. This quantum enhancement is manifest as a better scaling of estimator uncertainty with the number of particles into the interferometer, the primary resource in most quantum metrological schemes.


Most general interactions with a sample, however, comprises both Hamiltonian and dissipative parts in its dynamics. In the evolution of a probe through an interferometer, this amounts to the estimation of phase and loss parameters that characterise these two elements. This problem has important fundamental and technological implications, not only because all practical systems exhibit loss, but also because there are many situations in which the simultaneously estimating multiple parameters of several kinds is the objective. Examples include experiments where both dispersion and absorption profiles of a sample are sought with high accuracy using a single experimental setup, as well as Mueller polarimetric imaging~\cite{Pierangelo:11}. It is possible to estimate the parameters independently by preparing different optimal probe states and measurements for each parameter; however this is tantamount to preparing a different experimental setup for each parameter and requires a potentially demanding experimental reconfiguration. This is unsuitable for sensing time-varying samples, for instance.

These limitations can be overcome in principle by simultaneous estimation of both parameters, thus casting the problem in the framework of multiparameter estimation~\cite{Ballester04,ImaiFujiwara07,Genoni_08,Paris09,Chiuri_RVPIGMM11,Blandino_2012, Genoni12}. The main challenge in multiparameter quantum metrology arises from the non-attainability of the bound when the infinitesimal generators associated with the parameters do not commute \cite{Yuen_L73, HelstromKennedy, Belavkin76, fujiwara01}. Early studies on the limits of estimating a complex parameter~\cite{Yuen_L73, HelstromKennedy, Belavkin76,Genoni12} identified the right logarithmic derivative (RLD), which led to non-hermitian measurements, as attaining the most informative bounds. Interestingly, in our case, we find the symmetric logarithmic derivative (SLD) to play that role. In the general area of quantum multiparameter metrology, investigations into the role of entangled measurements in multiparameter estimation in specific problems such as qubit state estimation~\cite{Hayashi2008} have been undertaken. The estimation of Gaussian channels with Gaussian resources~\cite{Monras_2010,Monras_2011}, including cases when joint measurements employs asymptotically many copies of probes and channels has also been studied.


\begin{figure}[b*]
\includegraphics[viewport= 80 300 600 550, clip, width=\columnwidth]{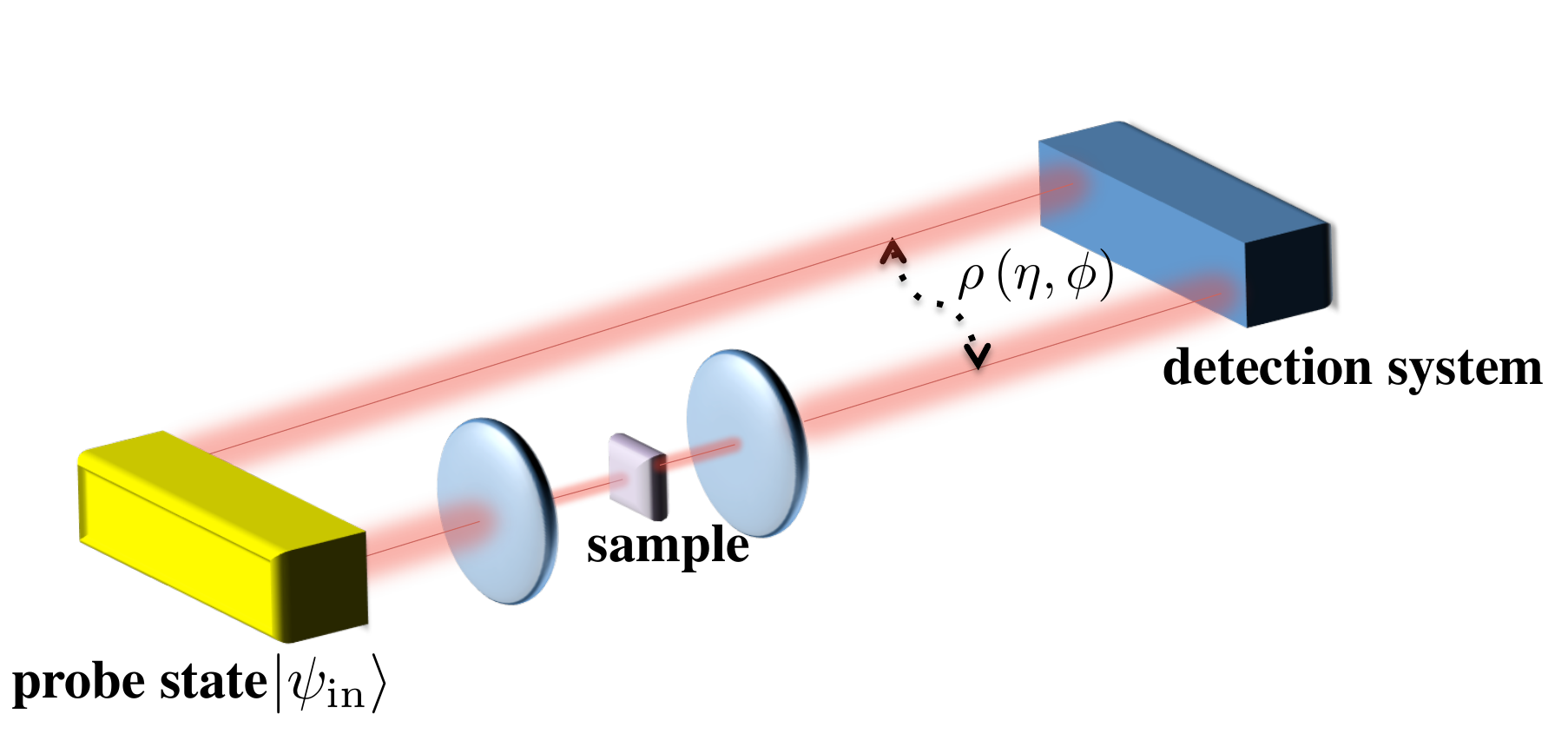}\\
  \caption{Schematic of an imaging system : A two-mode $n$-photon quantum probe state $\ket{\psi_{\rm{in}}}$ irradiates the sample and provides the reference arm of a two-port interferometer. The detection is chosen so to estimate simultaneously the phase shift $\phi$ (Hamiltonian dynamics) and the transmission $\eta$ (dissipative dynamics) from the state $\rho(\phi,\eta)$ resulting from the interaction of the probe with the sample.}
  \label{fig:WithLoss}
\end{figure}

In this paper, we investigate the limits of the precision with which phase and loss in an interferometer can be estimated simultaneously using a fixed number of photons. We derive a trade-off necessitated between the precision bounds in the simultaneous estimation of phase and loss by a given measurement strategy. We also identify the appropriate quantum probes for this scenario, providing the design of optimal two-mode quantum states that come closest to the ultimate quantum limit of simultaneous phase and loss estimation. These are found for fixed photon numbers by explicit numerical optimization.





\section{The framework}

The archetypal schema for quantum sensing is illustrated in Fig.~(\ref{fig:WithLoss}). An object, characterized by a set of parameters $\bm{\theta}=\left\{\theta_\nu \right\},$ is placed in one of the arms of a Mach-Zehnder interferometer. This extends the simple case in which a single phase shift contains the only relevant information, to a more realistic case, in which, for example, both the phase shift and the loss are important. The initial probe evolves upon propagation through the system, acquiring a form that depends on the parameter set. At the output, measurements of the probe state provide a multivariate probability distribution that captures changes in the state due to changes in the system parameter. The optimal probe state is the one which maximises the information that this distribution contains about the parameters.

For a single parameter, the optimal states is identified using a procedure which maximises the quantum Fisher information. This places a lower limit on the variance of the parameter's estimator. A quantum analogue to the score, or logarithmic derivative, is required to derive this bound. For single parameter estimation, the appropriate bound is provided by using the SLD. The logarithmic derivative quantifies the changes of the state with respect to the system parameters. The eigenbasis of the SLD provides the optimal measurement strategy which saturates this quantum limit on the variance of the estimator, the quantum Cram\'er-Rao bound~\cite{Braunstein_1994,Paris09}.

In the multivariate problem the estimator variance is promoted to a covariance matrix $\cv(\bm{\theta}),$ and is bounded by the inverse of the quantum Fisher information matrix through the quantum Cram\'er-Rao bound~\cite{Helstrom_1976}
\be
\label{eq:qii}
\cv(\bm{\theta}) \geq(M\I{\bm{\theta}})^{-1},
\ee
where $\I{\bm{\theta}}$ is the quantum Fisher information matrix associated with the evolved quantum state, and $M$ is the number of times the experiment is repeated. Clearly the precision of any estimate can be improved by repeating the experiment multiple times no matter what the input state, so that $M$ provides a purely classical advantage. Any quantum advantage is captured by $\I{\bm{\theta}}.$ As our interest is specifically in the potentially greater advantages permitted by using quantum estimation strategies, we will suppress the quantity $M$ in the following discussion.

Eq.~(\ref{eq:qii}) is in general a matrix inequality, though for the special case of single parameter estimation is given by the scalar Cram\'{e}r-Rao inequality $(\Delta \theta_{\nu})^2 \geq \I{\nu\nu}^{-1}$ in terms of the quantum Fisher information $\I{\nu\nu}$~\cite{Braunstein_1994}. For a single parameter $\theta_{\nu}$, the quantum Cram\`{e}r-Rao bound is always attainable, with the optimal measurement being given by the eigenvector of the symmetric logarithmic derivative (SLD) defined as~\cite{Braunstein_1994,Paris09}
\be
\label{eq:sld}
L_{\nu}\rho(\bm{\theta}) + \rho(\bm{\theta}) L_{\nu} = 2 \frac{\partial \rho(\bm{\theta})}{ \partial \theta_{\nu}},
\ee
which also defines the quantum Fisher information matrix as
\be\I{\mu\nu}=\tr{\rho(\bm{\theta})\frac{L_{\mu}L_{\nu}+L_{\nu}L_{\mu}}{2}}.
\ee
The precision in the estimate of the parameters is given by the saturation of the inequality in $(\Delta\bm{\theta})^2 = \sum_{\nu}(\Delta \theta_{\nu})^2 =\tr{\cv(\bm{\theta})} \geq \tr{(\I{\bm{\theta}})^{-1}}.$ The classical multiparameter information inequality can always be achieved asymptotically using unbiased estimators, but in general, the quantum matrix equality in Eq.~(\ref{eq:qii}) cannot be achieved~\cite{Helstrom_1976}.

\section{The Cram\'er-Rao bound for joint Phase and loss estimation}

The best probes states are the ones that maximize the quantum Fisher information. In our analysis, we focus on pure states of a fixed particle number. The most general pure states of a fixed number of photons are of the form
\be
\label{psi0}
\ket{\psi_{\rm{in}}} = \sum_{k=0}^n \alpha_k \ket{k,n-k}.
\ee
The same final state $\rho(\bm{\theta})$ is obtained independent of the order in which the phase accumulation and loss operator is applied~\cite{Demkowicz-Dobrzanski_2009} on the initial state in Eq.~(\ref{psi0}). Upon propagation through loss, the input state is transformed as a direct sum
\be
\rho(\bm{\theta}) = \bigoplus_{l=0}^n p_l \ket{\psi_l} \bra{\psi_l}.
\label{after}
\ee
where each term $\ket{\psi_l}$ is associated to the loss of $l$ photons; explicit expressions are provided in Appendix A.

These identify subspaces which remain orthogonal under infinitesimal translations. Therefore, the SLD is similarly block diagonal with (see Appendix A)
\ben
\label{onemodesymlog}
        & & L_{\nu} = \bigoplus_{l=0}^n L^l_{\nu}, \mbox{~~~~~with}  \\
L^l_{\nu} &=&\bigg((\partial_{\nu} \log p_{l})\ket{\psi_{l}}\bra{\psi_{l}}+ 2  \ket{\partial_{\nu} \psi_{l}}\bra{\psi_{l}}+2 \ket{\psi_l}\bra{\partial_{\nu} \psi_l} \bigg), \nonumber
\een
and $\ket{\partial_{\nu} \psi_l}$ is the partial derivative of $\ket{\psi_l}$ with respect to $\theta_\nu.$ Given this block diagonal structure, the elements $\I{\mu\nu}{}$ of the quantum Fisher information matrix are given by
\be
\label{QFI:onemode}
\I{\mu\nu}[\rho]= I_{\mu\nu}\left[\bm{p} \right] + \sum_{l=0}^{n}p_{l}\I{\mu\nu}\left[ \ket{\psi_l} \bra{\psi_l} \right],
\ee
where $\bm{p}$ is the vector of probabilities $\{p_l\}.$ The first term $I_{\mu\nu}[\bm{p}] = \sum_{l=0}^n p_{l} \, \partial_\mu \log p_{l}\partial_\nu \log p_{l}  $ is the classical Fisher information matrix of the probability distribution $\bm{p}.$ The second term has no classical analogue, and is interpreted as the quantum contribution to the information. It is the weighed quantum Fisher information of the constituent states $\I{\mu\nu}[ \proj{\psi_l}] = 4 \sum_{l=0}^n p_{l}(\Re P_{\mu\nu,l}),$ where we use the shorthand
\be
P_{\mu\nu,k} = \bra{\partial_{\mu} \psi_k} \Pi_k \ket{\partial_{\nu} \psi_k},
\ee
and the operator $\Pi_k = \mathbb{I} - \ket{\psi_k}\bra{\psi_k}$ projects onto the subspace orthogonal to $\ket{\psi_k}.$ For loss and phase estimation using states of the form Eq.~(\ref{psi0}), the quantum Fisher information matrix
\be
\mathcal{I}[\rho] = \left(
                \begin{array}{cc}
                  \I{\phi\phi} & \I{\phi\eta} \\
                  \I{\eta\phi} & \I{\eta\eta} \\
                \end{array}
              \right),
\ee
is such that $\I{\eta\phi}=\I{\phi\eta}=0.$

The covariance matrix for the phase shift $\phi$ (ranging between 0 and $2\pi$) and loss $\eta$ (ranging between $0$ for complete absorption and $1$ for complete transmission of the probe light) is diagonal when the quantum Cram\'er-Rao bound is saturated. In particular, the quantum Fisher information matrix is given by
\be
\label{eq:qfimatrix}
\mathcal{I}[\rho] = \left(
                \begin{array}{cc}
                  \I{\phi\phi} & 0 \\
                 0 & \I{\eta\eta} \\
                \end{array}
              \right),
\ee
where $\I{\phi\phi}, \I{\eta\eta}$ are the quantum Fisher information for the estimation of phase and loss respectively. Defining $\xi_{r,l} = \sum_{k=l}^n x_k b^k_l k^r, ~ \Xi_r  = \sum_{k=0}^n x_k k^r,$ the moments of the coefficients of $x_k=|\alpha_k|^2,$ $b^k_l$ a binomial factor~(see Appendix A), with
\be
\label{eq:qfietaphi}
\I{\phi\phi} = 4\left(\Xi_2 - \sum_{l=0}^n\frac{\xi^2_{1,l}}{\xi_{0,l}}\right), ~~ \I{\eta\eta} = \frac{\Xi_1}{\eta(1-\eta)},
\ee
it is easy to conclude the form of the optimal states for the estimation of loss and phase independently. In the absence of losses, $\eta = 1,$ $b^k_l = \delta_{l,0},$ $\I{\phi\phi}^{(\eta=1)} = 4\left(\Xi_2 - \Xi^2_1 \right),$ which is the variance of $x_k,$ maximised for a $n00n$ state, when $\I{\phi\phi}^{(\eta=1)} \sim n^2,$ as is well-known. In the lossy case, the best states for estimation of phase are arrived at by maximizing the general form of $\I{\phi\phi}$ \cite{Dorner_et_al_2009,Demkowicz-Dobrzanski_2009,Knysh_2011}. On the other hand, the best state for estimating the loss is the Fock state $\ket{n,0},$ in which case $\I{\eta\eta} \sim n$~\cite{Ji_WDFY08,Adesso_2009}.
For $\eta{=}0,1$ the quantum Fisher information for loss $\I{\eta\eta}$ diverges. This is also to be expected as in these cases all the photons are lost or all transmitted, so that the variance in the outcomes of measurements of particle number in these cases will be zero.

A necessary condition for saturating the multiparameter quantum Cram\'er-Rao bound is given by a vanishing expectation value of the commutator of the two SLDs~\cite{Guta2012}. In our case, this value is
\be
\label{eq:attainblock}
\tr{\rho [L_{\eta},L_{\phi}]} = i\sum_{l=0}^n p_l  \left( \Im \braket{\partial_{\mu} \psi_l}{\partial_{\nu} \psi_l} \right) = -i\frac{\I{\phi\phi}}{2\eta}.
\ee
This implies that the optimal measurements necessary to attain the quantum limits for the two parameters do not commute, and it is impossible to estimate the phase and loss using quantum probes with a fixed photon number.

In spite of the unattainability of the joint bound using fixed photon number states, we can aim for the most informative bound. This notion was developed to address quantum multiparameter estimation. However, unlike the case of estimating a complex amplitude parameter where the most informative bound is provided by a RLD~\cite{Genoni12}, for our problem is always the SLD.  We prove this in Appendix C. This shows that while our problem may appear similar to those studied earlier~\cite{Yuen_L73, HelstromKennedy, Genoni12}, the final solutions are quite different.

The origin of the trade-off between precisions in phase and loss estimation can be understood intuitively from the fact that the derivatives $\ket{\partial_\eta \psi_l}$ and  $\ket{\partial_\phi \psi_l}$ are the same up to an imaginary constant. Thus the local structure defines a two-dimensional Hilbert space spanned by $\ket{ \psi_l}$ and  $\ket{\partial_\phi \psi_l}$, and estimating the two parameters effectively corresponds to measuring along the axes given by $\ket{\psi_l}\pm\ket{\partial_\phi \psi_l}$ and $\ket{\psi_l}\pm i\ket{\partial_\phi \psi_l}$ \cite{fujiwara01}.

While this intuition gives a qualitative understanding of the origin of the trade-off, it provides no quantitative bounds on simultaneous estimation of phase and loss. We have quantified this trade-off in terms of Fisher information for a particular choice of measurement strategy. The left-hand side of Eq.~(\ref{eq:attainblock}) must equal zero for multiparameter estimation at the quantum Cram\'{e}r-Rao bound. In our case, the only instance when the quantum Cram\'er-Rao bound \textit{can} be saturated is when we learn nothing of the phase $\phi.$ Similarly, investing all of the available resources in estimating the loss parameter $\eta$ corresponds to using the Fock state $\ket{n,0}.$ Evidently, this state and the corresponding SLD has no sensitivity to the phase in the interferometer. More generally, no phase information can be extracted using a fixed photon number state as in Eq.~(\ref{psi0}) and measurements derived from the loss SLD.

\begin{figure}[!t]
\includegraphics[width=\columnwidth]{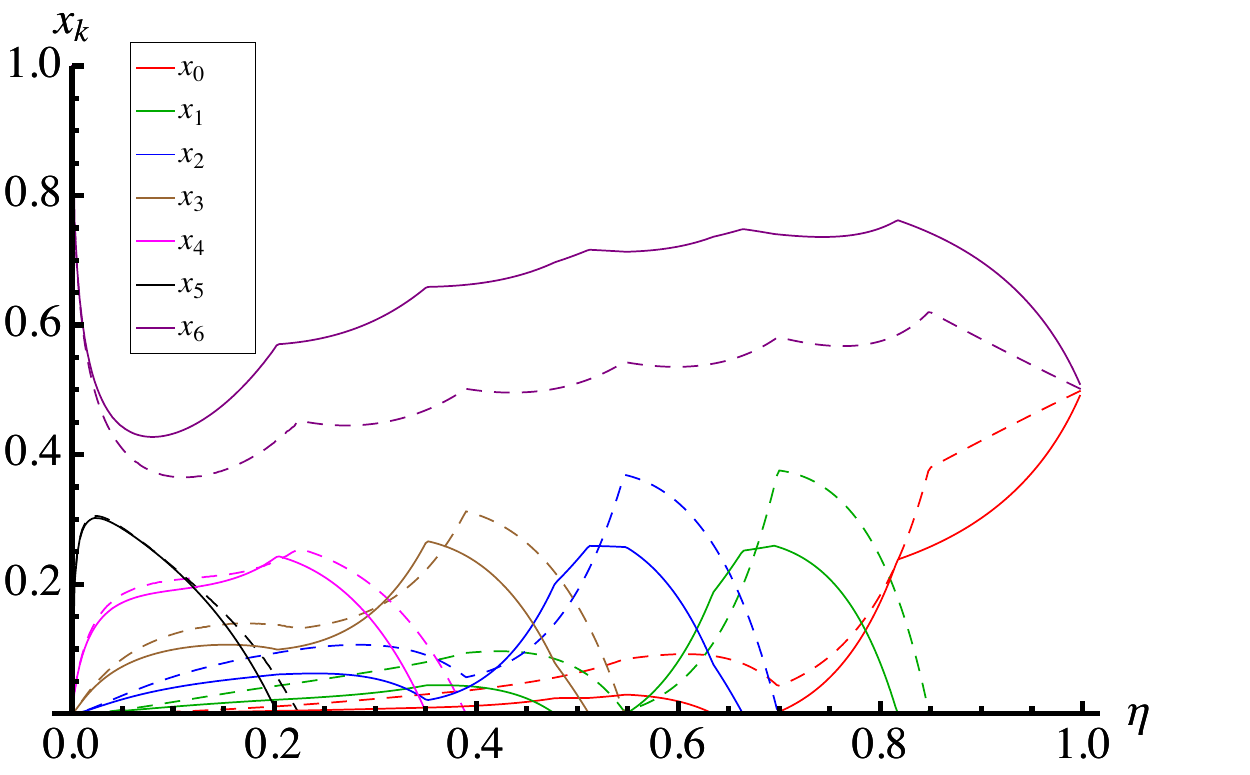}
\caption{[Color online] Coefficients of the optimal probe states of the form in Eq.~(\ref{psi0}) for $n=6$ obtained by numerical optimisation. The state resembles the optimal probes for phase estimation in the presence of losses~\cite{Dorner_et_al_2009}, and interpolates towards the Fock state $\ket{6,0},$ optimal state for estimating $\eta.$ Solid lines: simultaneous estimation of phase and loss. Dashed lines: estimation of only the phase in the presence of losses~\cite{Dorner_et_al_2009}.}
\label{fig:coef}
\end{figure}

The option we consider here is to use the SLD for phase estimation to estimate the loss parameter. This approach differs from what has been considered in \cite{Yuen_L73,HelstromKennedy}, since we do not need to implement a non-selfadjoint measurement operator based on the RLD. Our choice at least guarantees the quantum limit for one of the parameters, while any other measurement will be suboptimal for both. Using as projectors $\ket{\lambda},$ the eigenvectors of the phase SLD $L_\phi,$ to estimate both parameters, the measurement scheme results in a probability distribution $\{q_\lambda=\bra{\lambda}\rho\ket{\lambda}\}.$ Analysing this for information yields, as expected, values tight to the quantum bound for $\phi$, but a lower value of information about $\eta$~(see Appendix B)
\be
\label{eq:Ietaeta}
I_{\eta\eta}  = \I{\eta\eta} - \frac{1}{4\eta^2}\I{\phi\phi}.
\ee
This result brings to light interesting features in the joint estimation trade-off. As expected, the precision of loss estimation can only be enhanced at the cost diminished precision in estimating the phase, and vice-versa. In spite of this tradeoff, there is an asymmetry between the two parameters. Estimating loss at the quantum limit requires a complete sacrifice in estimating phase, but phase estimation at the quantum limit leaves us with some information about loss. Thus, quantum mechanics limits the estimation of Hamiltonian and dissipative parameters at the quantum limit differently, given by Eqns.~(\ref{eq:qfietaphi}) and (\ref{eq:Ietaeta}).


\begin{figure}[!t]
\includegraphics[width=\columnwidth]{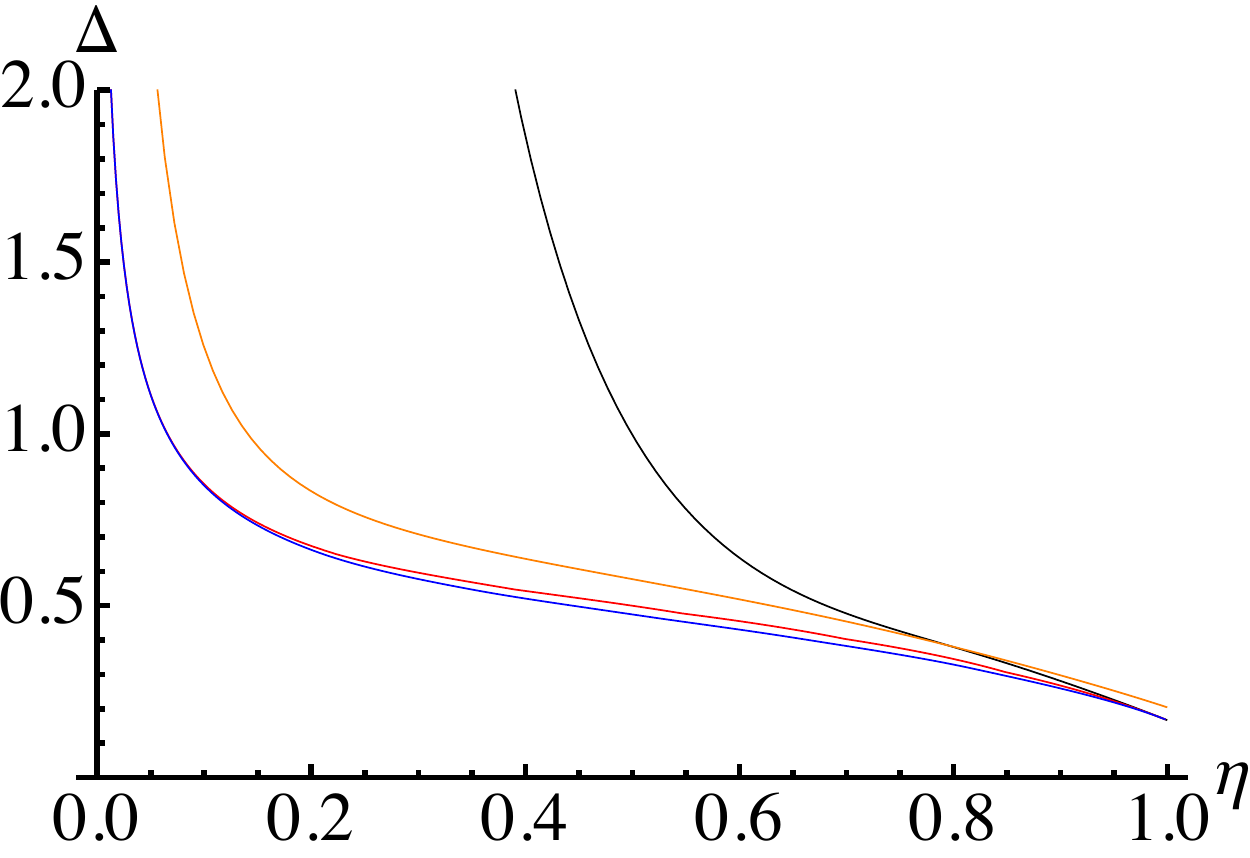}
\caption{[Color online] Precisions attainable in the combined estimation of loss and phase with different quantum probe states for $n=6.$ $n00n$ states (black), Holland-Burnett states (orange), optimal lossy phase estimation states (Red), and our optimal states for simultaneous phase and loss estimation(Blue). $\eta$ is the loss and $\Delta=\sqrt{(\I{\phi\phi})^{-1}+(I_{\eta\eta})^{-1}}.$}
\label{fig:comp}
\end{figure}

\section{Designing the optimal probe states}
In addition to the conceptual understanding of quantum trade-offs and limitations, the practical challenge in any multiparameter estimation scenario such as imaging in Fig.~(\ref{fig:WithLoss}) lies in identifying the quantum probes that maximize the precisions of both the parameters simultaneously to the best possible extent. For a single parameter, this is tantamount to maximizing the quantum Fisher information. For multiparameter quantum metrology, the sum of the variances of all the parameters is bounded from below by the trace of the inverse of the quantum Fisher information matrix. In our case of a diagonal Fisher information matrix, the optimal probes are then given by the minimization of the sum of the reciprocals of $\I{\phi\phi}$ and $I_{\eta\eta}$ given by Eqns.~(\ref{eq:qfietaphi}) and (\ref{eq:Ietaeta}). We have performed simulations for particle numbers up to  200. Here we present as an example the result for $n=6$ in Fig.~(\ref{fig:coef}). As can be seen, higher weight is shifted to the lossy arm to aid the estimation of  loss at an enhanced precision. This appears as a general feature for arbitrary values of $n$; examples are reported in Appendix D. As quantum states with an increasing number of photons are engineered~\cite{8photon}, the experimental generation of our probe states and their application in multiparameter sensing may be possible in the foreseeable future.

\begin{figure}[!t]
\includegraphics[width=\columnwidth]{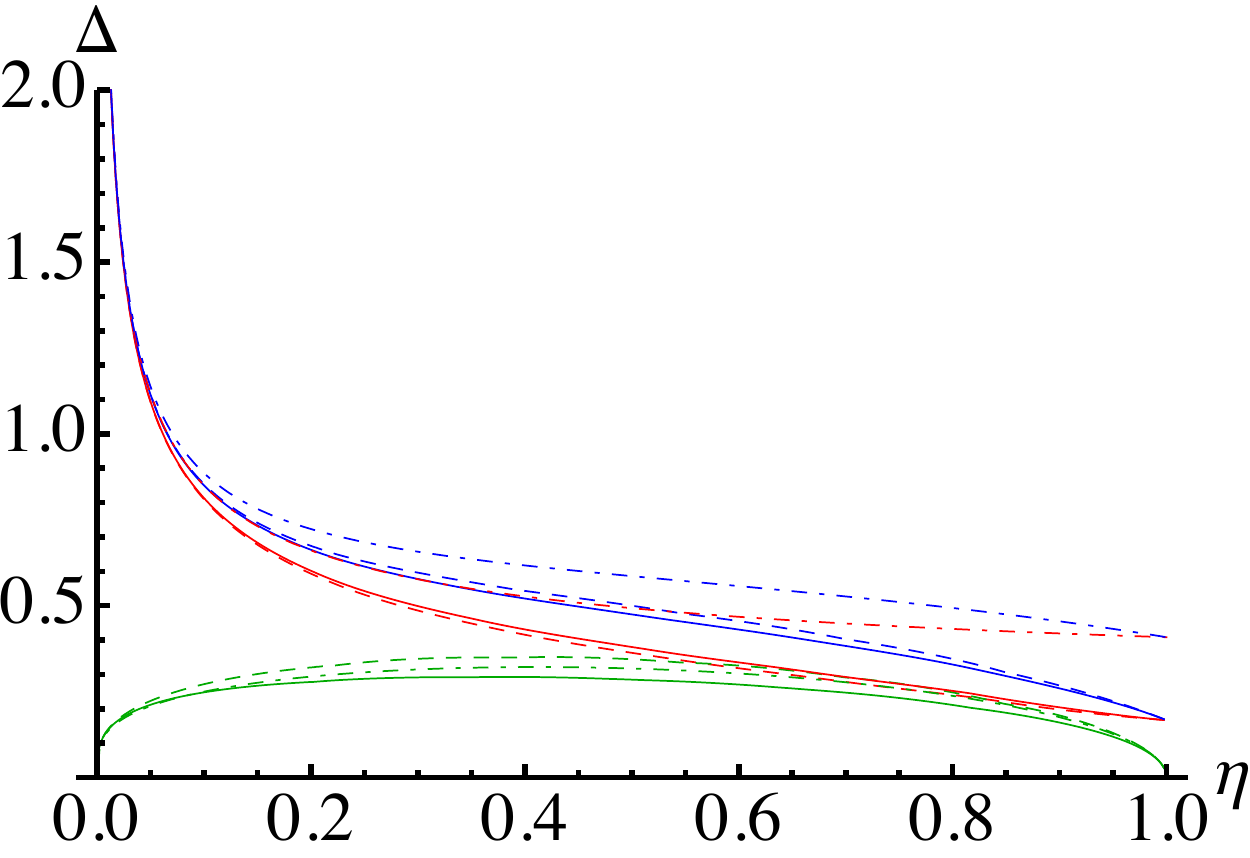}
\caption{[Color online] The contributions to the combined precision $\Delta\bm{\theta}$ (Blue) separated into the phase $\Delta\phi$ (Red) and loss $\Delta\eta$ (Green) parts. The solid and dashed lines represents the precisions for the optimal state for simultaneous phase and loss, and phase estimation, given by the solid lines and dashed lines respectively in Fig.~(\ref{fig:coef}). All plots are for $n=6.$ For comparison, we also report the precisions attainable with a coherent state probe $\ket{\alpha}$ with amplitude $|\alpha|^2{=}6$ as the dash-dotted line in terms of the standard interferometric limit \cite{Dorner_et_al_2009}. All the curves are independent of $\phi$ and $\Delta\phi = (\I{\phi\phi})^{-1/2},\Delta\eta = (I_{\eta\eta})^{-1/2},$ and $(\Delta\theta)^2 = (\Delta\phi)^2+(\Delta\eta)^2.$}
\label{fig:cont}
\end{figure}

In Fig.~(\ref{fig:comp}), we compare the performance of the optimal probe states with other commonly used states such as the $n00n$ states~\cite{Lee_Kok_Dowling_2002}, Holland-Burnett states~\cite{Holland_1993}, and the optimal states for phase estimation in the presence of single mode losses. The similarity of performance between our optimal state and the optimal state for lossy phase estimation is not unexpected since the dominant contribution to $\Delta\bm{\theta}$ is from the uncertainty in estimating $\phi.$

This is demonstrated in Fig.~(\ref{fig:cont}). The gap between the solid and dashed blue lines also shows the overall improvements in the combined precision of phase and loss estimation when using our optimal states over the entire range of $\eta$. As is evident, we gain more in the estimation of $\eta$ than we loose in the estimation of $\phi.$ {\color{black} A comparison with the performance with a coherent state of similar intensity, as captured by the standard interferometric limit \cite{Dorner_et_al_2009}, shows that an advantage can be obtained mostly for the phase parameter.}  Fig.~(\ref{fig:comp}) also reveals that the performance of a Holland-Burnett probe state is close to optimal. This provides a feasible route to obtaining close to the optimal precisions in the laboratory~\cite{Datta_2011}.

\begin{figure}[!t]
\includegraphics[width=\columnwidth]{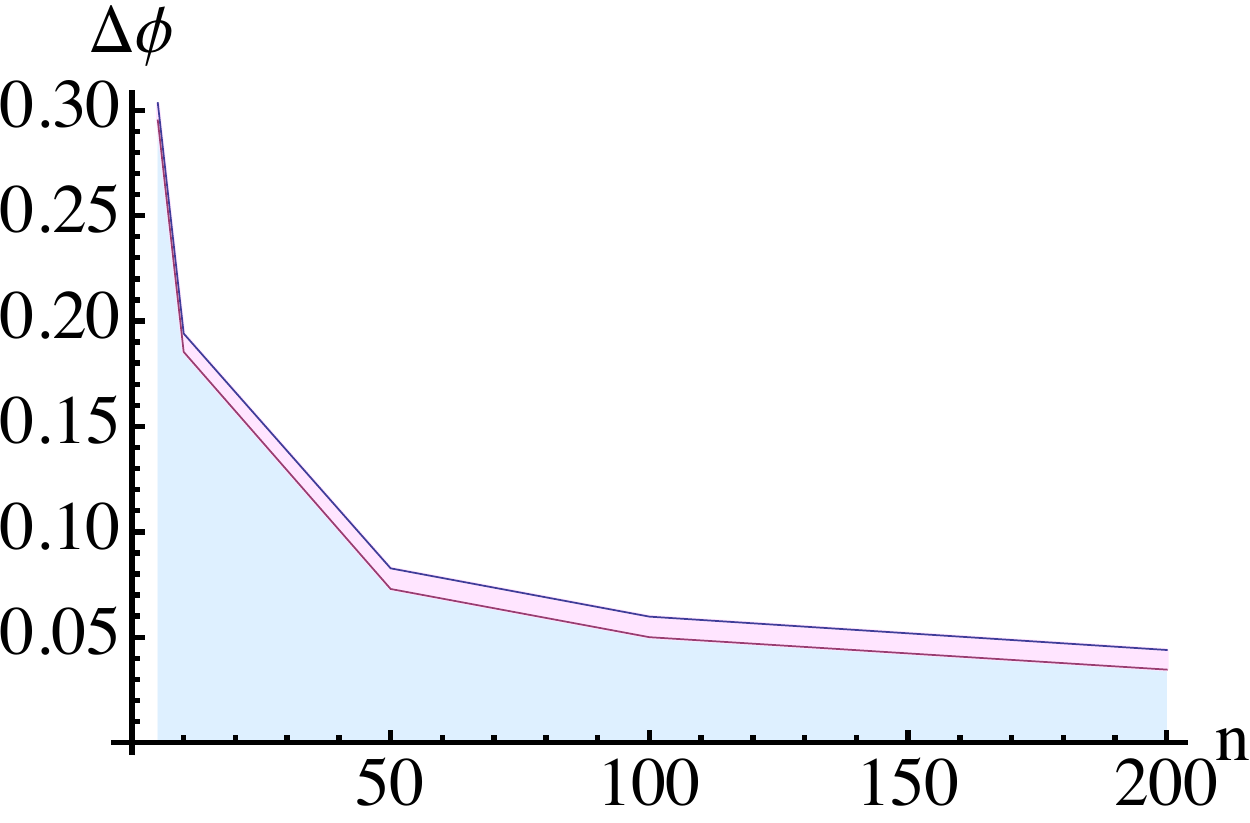}
\caption{[Color online] Phase uncertainty $\Delta \phi$ for the optimal phase estimation scheme (blue) and for the multiparameter scheme (magenta) for photon number $n{=}5, 10, 50, 100, 500$, and $\eta{=}0.9$. }
\label{fig:comparephi}
\end{figure}

Optimising on multiple parameter estimation has to come at the price of reduced phase sensitivity. This is quantified in Fig.~(\ref{fig:comparephi}), in which we compare the uncertainty for phase only for our states and the optimal states for phase estimation for increasing $n$ at low loss, where we expect quantum advantage to be more relevant. Although the added uncertainty grows with the intensity, this remains under 20\%  at moderate photon numbers.

\section{Conclusions}

The quantum Fisher information matrix for phase and loss estimation is diagonal. This is true for any estimation problem involving a pair of parameters generating Hamiltonian and dissipative dynamics independently, as the expectation value of the overlap between the optimal measurements for such parameters is always completely imaginary, the real part of which is zero. This is true even for number non-conserving probe states. The matrix inequality that defines the quantum Cram\'er-Rao bound does not, however, take into account the non-commutativity of the optimal quantum measurements, which prevents it from being saturated. This makes multiparameter quantum metrology nontrivial, and fundamentally different not only from classical multiparameter estimation, but also quantum single parameter estimation, as in both these cases the Cram\'er-Rao bound can always be saturated.

Our main advance is to quantify how quantum mechanics limits the simultaneous estimation of phase and loss parameters at the quantum limit when using a fixed number of photons. In particular, we have illustrated that the estimation of a pair of parameters describing the Hamiltonian and the dissipative evolution of a probe state in an interferometric sensor cannot give both parameters at the simultaneous quantum limit of each.

Multiparameter estimation is relevant to a broad range of sensors, and opens the door to more complex quantum imaging devices, in which multiple parameters related to the object configuration are sought. The best possible strategy imposes a trade-off between the attainable precisions of the parameters. This trade-off forces an optimization strategy to determine the form of best quantum probes for multiparameter quantum metrology at the ultimate attainable limit.


{\bf Acknowledgements} We thank G.~Durkin and the members of his group for elucidating the origin of the trade-off in the noncommutativity of the infinitesimal generators. We acknowledge stimulating and informative discussions with M.~Guta, J.~Nunn, X.~Jin, L.~Zhang and M.~Genoni. This work was funded in part by EPSRC (Grant No. EP/H03031X/1), U.S. EOARD (Grant No. 093020), EU Integrated Project QESSENCE. MB is supported by a FASTQUAST ITN Marie Curie fellowship. IAW acknowledges support from the Royal Society.

\bibliography{MastersRefs}

\appendix

\section{Fixed-photon-number states}
\label{sec:fixedphoton}

We use pure quantum probe states with a fixed resource, in our case photon number $n$, to estimate simultaneously the loss and phase in the two-mode interferometer in Fig.~(1).
\be
\ket{\psi_{\rm{in}}} = \sum_{k=0}^n \alpha_k \ket{k,n-k}
\ee
The loss is modeled as a beam splitter of transmissivity $\eta$ on the same mode that imprints the phase. At this juncture, the mode mixes with the vacuum resulting in a loss mode $\ket{l}_L$. The loss transforms the basis vectors as $\ket{k,n-k} \mapsto \sum_{l=0}^{k} \sqrt{b_{l}^k } \ket{k-l,n-k} \otimes \ket{l}_L,$ where the coefficients $b_l^k$ are given by $b_{l}^k= \binom{k}{l} \eta^{k-l}(1-\eta)^{l}.$ Including the phase accumulation, the final state is given by
\be
\ket{\psi} = \sum_{k=0}^n \alpha_k e^{i k \phi} \sum_{l=0}^{k} \sqrt{b_{l}^k } \ket{k-l,n-k} \otimes \ket{l}_L.
\ee
The state of the loss mode cannot be measured, so to reflect this loss of information, we sum over all possible states by tracing out the mode $L$. This leads to the evolved probe state $\rho = \sum_{l=0}^n  p_{l} \ket{\psi_{l}} \bra{\psi_{l}},$ where the normalised states $\ket{\psi_l}$ are given by
\be
\label{generalket}
\ket{\psi_l} = \frac{1}{\sqrt{p_l}} \sum_{k=l}^{n-l} \alpha_k \sqrt{b^k_l} e^{i k \phi} \ket{k-l,n-k}.
\ee
The orthogonality of the states $\braket{\psi_{l}}{\psi_{l'}}=\delta_{ll'},$ allows the density matrix to be written as a direct sum as ~\cite{Demkowicz-Dobrzanski_2009}
\be
\label{1modestate}
\rho = \bigoplus_{l=0}^n p_l \rho_l = \bigoplus_{l=0}^n p_l \ket{\psi_l} \bra{\psi_l}.
\ee

\textbf{Multiparameter estimation strategy:} Our analysis makes use of a helpful properties of the evolved state. Noting that state vectors $\ket{\psi_l}$ remain orthogonal under infinitesimal translations in either parameter of the form $\theta_\kappa \mapsto \theta_\kappa + \mathrm{d}\theta_\kappa$, (i.e. $( i \neq j ) \braket{\partial_\kappa \psi_i}{\psi_j} =0$\PC{)}, then the derivative $\partial_\kappa \rho$ and the symmetric logarithmic derivative (SLD) $L_\kappa$ decompose into diagonal blocks $\partial_\kappa \left( p_l \rho_{l} \right)$ and $L_{l,\kappa}$ supported on the same orthogonal subspaces as the pure states $\rho_{l}$. This block diagonal form means that eigenvalues of $L_\kappa$ are simply the eigenvalues of the blocks $L_{l,\kappa}$. Each block has two non-zero eigenvalues:
\begin{equation}
\label{eigval}
\lambda_{l,\kappa}^{\pm} = \frac{1}{2} \left( \partial_\kappa \log p_l \pm \sqrt{ (\partial_\kappa \log p_l)^2 + 16 P_{l,\kappa\kappa} } \right)
\end{equation}
each with the corresponding eigenvector
\begin{equation}
\label{eigvec}
\ket{ \lambda_{l,\kappa}^{\pm} } = \frac{  \lambda_{l,\kappa}^{\pm}  \ket{\psi_l} + 2 \mathbb{P}_l \ket{\partial_\kappa \psi_l} }{\sqrt{   \left(\lambda_{l,\kappa}^{\pm}\right)^2  + 4 P_{l,\kappa\kappa}  }}
\end{equation}
where $P_{l,\mu\nu} = \bra{\partial_\mu \psi_l} \Pi_l \ket{\partial_\nu \psi_l}$, and the operator $\Pi_l = \mathbb{I} - \ket{\psi_l}\bra{\psi_l}$ projects into the space perpendicular to $\ket{\psi_l}$. If this strategy is employed to saturate the Cram\'{e}r-Rao bound for estimation of the parameter $\theta_\kappa$, measurement records one of $2n+2$ final states $\ket{ \lambda_{l,\kappa}^{\pm} }$ with probability $q_l^{\pm} = \bra{ \lambda_{l,\kappa}^{\pm} } \rho \ket{ \lambda_{l,\kappa}^{\pm} }$. This distribution will also contain information $I_{\mu\nu}\left(\theta_\kappa\right)$ on other parameters. This is given by the classical Fisher information of the distribution $\{ q_l^+, q_l^- \}_{l=1}^n$ and will generally deviate from the quantum bound $\I{\mu\nu}$,
\be
I_{\mu\nu} \left( \theta_\kappa \right) =  I_{\mu\nu} \left[  \{ q_l^+, q_l^- \}_{l=0}^n \right]
\ee
In our case it is found that the distribution $q_l^{\pm} = \bra{ \lambda_{l,\eta}^{\pm} } \rho \ket{ \lambda_{l,\eta}^{\pm} }$ contains no information on $\phi.$ So it is not a useful multiparameter estimation strategy, whereas the alternative distribution $q_l^{\pm} = \bra{ \lambda_{l,\phi}^{\pm} } \rho \ket{ \lambda_{l,\phi}^{\pm} } = \frac{1}{2} p_l$ is. Thus
\be
I_{\eta\eta} \left( \phi \right) =  I_{\eta\eta} \left[  \{ \tfrac{1}{2}p_l, \tfrac{1}{2}p_l \}_{l=1}^n \right] = I_{\eta\eta}[ \bm{p}]
\ee

\section{Derivation of the quantum Fisher information matrix}

The state is given by Eqns.(\ref{generalket}) and \eqref{1modestate}. The Fisher information matrices for states of this form are given by Eq.~(6) in the main text. In what follows,
\be
\xi_{r,l} = \sum_{k=l}^n x_k b^k_l k^r, ~~\Xi_k = \sum_{l=0}^n \xi_{r,l}  = \sum_{k=0}^n x_k k^r,
\ee
are the moments of the coefficients of $x_k = |\alpha_k|^2,$ with the constraint  $\sum_{k=0}^n x_k =1,$ and $b^k_l$ is the binomial factor defined in the last section. We begin by reexpressing the probabilities
\begin{equation}
\label{app:pl}
p_{l} = p_{l} \braket{\psi_{l}}{\psi_{l}} = \sum_{k=l}^{n} \absq{\alpha_k} b_{l}^k = \sum_{k=l}^{n} x_k b_{l}^k = \xi_{0,l}.
\end{equation}
Evaluating the relevant inner products gives
\begin{subequations}
\label{app:innerprod}
\begin{equation}
\braket{\psi_{l}}{\partial_\phi \psi_{l}} = \i \sum_{k=l}^n \frac{k \absq{\alpha_k} b_l^k}{p_{l}}= \i \frac{\xi_{1,l}}{\xi_{0,l}},
\end{equation}
\begin{equation}
\braket{\partial_\phi \psi_{l}}{\partial_\phi \psi_{l}}= \sum_{k=l}^n \frac{k^2 \absq{\alpha_k} b_l^k}{p_{l0}} =  \frac{\xi_{2,l}}{\xi_{0,l}},
\end{equation}
\begin{eqnarray}
\nonumber
\braket{\psi_{l}}{\partial_\eta \psi_{l}}
&=& \sum_{k=l}^n \alpha_k^{\ast} e^{- i k \phi} \sqrt{\frac{b_l^k}{p_{l}}} \partial_\eta \left( \alpha_k e^{ i k \phi} \sqrt{\frac{b_l^k}{p_{l}}} \right) =  \\
&=& \frac{1}{2}  \partial_\eta \left( \frac{1}{\xi_{0,l}}  \sum_{k=l}^n x_k b_l^k \right) =  0.
\end{eqnarray}
This reflects the geometry of the estimation problem, wherein the $\eta$ derivative is orthogonal to the initial state,
\begin{eqnarray}
\nonumber
\braket{\partial_\eta \psi_{l}}{\partial_\eta \psi_{l}}&=&\sum_{k=l}^n \partial_\eta \left(\frac{\alpha_k^{\ast} e^{- i k \phi} \sqrt{b_l^k}}{\sqrt{p_{l}}} \right) \partial_\eta \left( \frac{\alpha_k e^{i k \phi} \sqrt{b_l^k }}{\sqrt{p_{l}}} \right) \nonumber \\
&=& \sum_{k=l}^n \absq{\alpha_k} \left( \partial_\eta \sqrt{\frac{b_l^k}{p_l}} \right)^2 \!\!=\! \frac{\xi_{2,l} \xi_{0,l} - \xi_{1,l}^2}{4 \eta^2 \xi_{0,l}^2},
\end{eqnarray}
\begin{eqnarray}
\braket{\partial_\phi \psi_l}{\partial_\eta \psi_l}
&=& - i \sum_{k=l}^n k \alpha_k^{\ast} e^{- i k \phi} \sqrt{\frac{b_l^k}{p_l}} \partial_\eta \left( \alpha_k e^{i k \phi} \sqrt{\frac{b_l^k}{p_l}} \right) \nonumber \\
&=&  -i \frac{ \xi_{2,l} \xi_{0,l} - \xi_{1,l}^2}{2 \eta \xi_{0,l}^2}.
\end{eqnarray}
\end{subequations}

Writing these in the shorthand $P_{k,\mu\nu} = \braket{\partial_\mu \psi_k}{\partial_\nu \psi_k}-\braket{\partial_\mu \psi_k}{\psi_k}\braket{\psi_k}{\partial_\nu \psi_k}$ introduced in the last section, we find
\begin{subequations}
\label{app:Plmunu}
\begin{eqnarray}
P_{l,\phi\phi} &=& \frac{\xi_{2,l} \xi_{0,l} - \xi_{1,l}^2}{\xi_{0,l}^2},~
P_{l,\eta\eta} = \frac{ \xi_{2,l} \xi_{0,l} - \xi_{1,l}^2}{4 \eta^2 \xi_{0,l}^2} \\
P_{l,\phi\eta} &=& -i\frac{ \xi_{2,l} \xi_{0,l} - \xi_{1,l}^2}{2\eta \xi_{0,l}^2}.
\end{eqnarray}
\end{subequations}
Also required is the Fisher information of the distribution $\{ p_l \}$, given by the terms $\avg{\partial_\mu \log p_l \partial_\nu \log p_l}$. In order to evaluate these one needs to use some properties of $\xi_{k,l},$ including
\begin{subequations}
\begin{eqnarray}
\partial_\eta b_l^k &=& b_l^k \left( \frac{k}{\eta}-\frac{l}{\eta(1-\eta)} \right) \\
\partial_\eta \xi_{k,l} &=&  \left( \frac{\xi_{k+1,l}}{\eta} - \frac{l \xi_{k,l}}{\eta(1-\eta)} \right) \\
\sum_{l=0}^n \xi_{k,l} &=& \Xi_{k}, ~~~\sum_{l=0}^n \frac{l \xi_{k,l}}{1-\eta} = \Xi_{k+1} \\
\sum_{l=0}^n \frac{l^2 \xi_{k,l}}{(1-\eta)^2} &=& \Xi_{k+2} + \frac{\eta}{1-\eta} \Xi_{k+1}
\end{eqnarray}
\end{subequations}
These relations can be used to evaluate
\begin{subequations}
\begin{eqnarray}
\nonumber
\avg{\partial_\eta \log p_l \, \partial_\eta \log p_l} &=& \sum_{l=0}^n \frac{1}{p_l} \left( \partial_\eta p_l \right)^2 = \sum_{l=0}^n \frac{1}{\xi_{0,l}} \left( \partial_\eta \xi_{0,l} \right)^2 \\ \nonumber
&=& \frac{1}{\eta(1-\eta)}\Xi_1 - \frac{1}{\eta^2} \left(\Xi_2- \sum_{l=0}^n \frac{\xi_{1,l}^2}{\xi_{0,l}} \right).
\end{eqnarray}
\end{subequations}
The other products are much simpler and follow trivially from $\partial_\phi p_l = 0.$ That is, $\avg{\partial_\phi \log p_l \, \partial_\phi \log p_l} = 0, \avg{\partial_\phi \log p_l \, \partial_\eta \log p_l} =0.$ Although derived for the loss and the phase parameters here, this general feature for any pair of parameters corresponding to Hamiltonian and dissipative dynamics.

Combining these results, the quantum Fisher information matrix elements from Eq.~(8) in the main text are thus given by
\begin{subequations}
\label{app:IQF}
\begin{eqnarray}
\label{appndQFIphiphi}
\I{\phi\phi} &=& \sum_{l=0}^n p_l \left((\partial_\phi \log p_l)^2 + 4 \Re P_{l,\phi \phi} \right) \nonumber \\
    &=& 4 \sum_{l=0}^n \frac{1}{\xi_{0,l}} \left(\xi_{2,l} \xi_{0,l} - \xi_{1,l}^2 \right)
    = 4 \left( \Xi_2 - \sum_{l=0}^n \frac{\xi_{1,l}^2}{\xi_{0,l}}\right) \\
\label{appndQFIetaeta}
\I{\eta\eta} &=& \sum_{l=0}^n p_l \left((\partial_\eta \log p_l)^2 + 4 \Re P_{l,\eta \eta} \right) \nonumber \\
 &=& \frac{\Xi_1}{\eta(1-\eta)} - \frac{1}{\eta^2} \left(\Xi_2- \sum_{l=0}^n \frac{\xi_{1,l}^2}{\xi_{0,l}} \right)
 + \sum_{l=0}^n \frac{\left( \xi_{2,l} \xi_{0,l} - \xi_{1,l}^2\right)}{\eta^2 \xi_{0,l}} \nonumber \\
\label{appndQFIetaeta}
&=& \frac{1}{\eta(1-\eta)}\Xi_1 \\
\I{\eta\phi} &=& \I{\phi\eta} = 0
\end{eqnarray}
\end{subequations}

When evaluating the Fisher information matrix elements using the optimal measurements for $\phi,$ we obtain the same forms for $\I{}$ except $\I{\eta\eta}$ where the useful cancelation which occurs in Eq.~\eqref{appndQFIetaeta} no longer occurs.
\begin{subequations}
\label{app:IFphi}
\begin{eqnarray}
\nonumber
I^{L_{\phi}}_{\phi\phi} &=& 4 \sum_{l=0}^n \frac{1}{\xi_{0,l}} \left(\xi_{2,l} \xi_{0,l} - \xi_{1,l}^2 \right) = 4 \left( \Xi_2 - \sum_{l=0}^n \frac{\xi_{1,l}^2}{\xi_{0,l}}\right)
\\
I^{L_{\phi}}_{\eta\eta}(\phi) &=& \frac{1}{\eta(1-\eta)}\Xi_1 - \frac{1}{\eta^2} \left(\Xi_2- \sum_{l=0}^n \frac{\xi_{1,l}^2}{\xi_{0,l}} \right) \\
I^{L_{\phi}}_{\eta\phi}(\phi) &=& I^{L_{\phi}}_{\phi\eta}(\phi) = 0
\end{eqnarray}
\end{subequations}
Also of interest is estimation using the optimal measurements for $\eta$. One finds that since $P_{l,\eta\phi}$ purely imaginary, and $p_l$ carries no $\phi$ dependence, so $I^{L_{\eta}}_{\phi\phi}=I^{L_{\eta}}_{\eta\phi} = I^{L_{\eta}}_{\phi\eta}=0,$ and
\be
\label{app:IFeta}
I^{L_{\eta}}_{\eta\eta}= \frac{\Xi_1}{\eta(1-\eta)}.
\ee
A final result that is of some significance is the condition for commutativity of SLDs~\cite{Guta2012}. In terms of the shorthand introduced in the last section, it is
\begin{equation}
\tr{\rho \left(L_\mu L_\nu - L_\nu L_\mu \right)}=4i \sum_{l=0}^n p_l \Im P_{l,\mu \nu},
\end{equation}
which for the case of simultaneous estimation of $\eta$ and $\phi$, using the results of~\eqref{app:pl},~\eqref{app:innerprod} and~\eqref{app:Plmunu}, is
\ben
\nonumber
\tr{\rho \left(L_\eta L_\phi - L_\phi L_\eta \right)} &=& 4 \sum_{l=0}^n \frac{1}{2 \eta \xi_{0,l}} \left(\xi_{2,l} \xi_{0,l} - \xi_{1,l}^2\right) \nonumber \\
&=& -i\frac{2}{\eta} \left(\Xi_2- \sum_{l=0}^n \frac{\xi_{1,l}^2}{\xi_{0,l}} \right).
\een

\section{Symmetric and right logarithmic derivative}
\label{app:sldVSrld}

We present a proof that the SLD provides the most informative quantum Cram\'{e}r-Rao bound for our problem. To decide on a choice of a logarithmic derivative, we consider a family of logarithmic derivatives $L_\mu$, parameterised by a measure $m.$ They are defined implicitly by
\be
\label{gld}
\frac{\partial \rho(\bm{\theta})}{ \partial \theta_{\mu}} \equiv \dpr = \int_0^1 \rho^t L_\mu \rho^{1-t} \mathrm{d}m(t)
\ee
To explain this choice, first a corresponding inner product is defined on the space of operators
\be
\left( A , B \right) = \tr{\int_0^1 \rho^t A \rho^{1-t} B^{\dagger} \mathrm{d}m(t) }
\ee
Then, for an unbiased estimator $A_\mu$ of the parameter $\theta_\mu$, which satisfies $\tr{\rho A_\mu} = \theta_\mu$, or
\be
\label{optest}
\partial_\mu \tr{\rho A_\mu} = \left( A_\mu , L_\mu \right ) = 1,
\ee
a Cram\'er Rao bound can be defined via the Cauchy-Schwarz inequality as
\be
\left( A_\mu, A_\mu \right) \geq \frac{1}{\left( L_\mu, L_\mu \right)}.
\ee
Hence our choice of logarithmic derivative encapsulates all $L_\mu$ which satisfy equation \eqref{optest}, and the quantum Fisher information matrix is given by
\be
\IF = \left( L_\mu, L_\nu \right) = \tr{\int_0^1 \rho^t L_\mu \rho^{1-t} L_\nu^{\dagger} \mathrm{d}m(t) }.
\ee
The measure $m(t)$ is the one which gives the most restrictive bound, maximising the error on the estimator. To proceed, the state studied in this paper is of the form $\rho = \bigoplus_{l=0}^n p_l \rho_l$ for pure states $\rho_l$, and the logarithmic derivative and Fisher information matrix decomposes similarly,
\be
L_\mu = \bigoplus_{l=0}^n \lmi
\ee
Thus the $\lmi$ can be solved for individually. For this we use that $\rho_l$ is pure and hence satisfies $\rho_l^t=\rho_l$ for $t>0$, and $\rho_l^t= \mathbb{I}$ for $t=0$, thus the integrands loses their dependence on $t$ and equation \eqref{gld} becomes
\be
\label{gldi}
\begin{array}{rcl}
\partial_\mu\ri &=& \int_0^1 \ri^t \lmi \ri^{1-t} \mathrm{d}m(t) \\
&&\\
&=& M_0 \lmi \ri + M_1 \ri \lmi + M_2 \ri \lmi \ri
\end{array}
\ee
with $M_0,M_1,M_2 \in \left[ 0,1 \right]$ and $M_0+M_1+M_2=1$, and are related to the measure $m(t)$ by
\be
M_0 = \lim_{\epsilon \rightarrow 0} \int_0^\epsilon \mathrm{d}m(t), \,\,\,\,\,\, M_1 = \lim_{\epsilon' \rightarrow 0} \int_{1-\epsilon'}^1 \mathrm{d}m(t).
\ee
By solving Eq.~\eqref{gldi} one finds, using $\rho_l = \pket \pbra$ and $\Pi_i = \mathbb{I} - \pket \pbra$, that
\be
\lmi = \dlpl \pket \pbra + \frac{1}{M_0} \Pi_l \dpket \pbra + \frac{1}{M_1} \pket \dpbra \Pi_l.
\ee
This can be used to evaluate the Fisher information
\be
\IF = \mathcal{I}_{\mu\nu}[\bm{p}] + \sum_{l=0}^n p_l \mathcal{I}_{\mu\nu}\left[\rho_l\right]
\ee
as
\be
\IF = I_{\mu\nu}[\bm{p}]+ \sum_{l=0}^n p_l \left( \frac{P_{l,\mu\nu}}{M_0}+\frac{ P_{l,\nu\mu}}{M_1} \right)
\ee
where $P_{l,\mu\nu} = \bra{\partial_\mu \psi_l} \Pi_l \ket{\partial_\nu \psi_l}$ is closely related to the single parameter quantum Fisher information given by $\mathcal{I}_{\mu\mu}\left[\rho\right]=P_{l,\mu\mu}$. It is generally complex for $\mu\neq\nu$, and $I_{\mu\nu}$ is the classical Fisher information of the probability distribution $\bm{p}$.

For our problem for phase ($\phi$) and loss ($\eta$) estimation using a fixed number of photons, we obtain
\ben
\mathcal{I}_{\phi\eta}\left[\rho\right] &=& \left(
\begin{array}{cc}
0 & 0 \\
0 & I_{\eta\eta} \\
\end{array}
\right) \\
&+&
\frac{ \I{\phi\phi}}{16 \eta^2 M_0 M_1}
\left(
\begin{array}{cc}
4 \eta^2 (M_0+M_1) & 2 i \eta (M_1-M_0) \\
2 i \eta (M_0-M_1) & (M_0+M_1)  \\
\end{array}
\right).\nonumber
\een
The first matrix has only the $\eta\eta$ element since $\bm{p}$ is independent of $\phi.$ The imaginary terms in the second is to be expected, and is nullified when the SLD is implemented, which we next show is the unique solution. From the above equation, it can be shown that $\vec{u}^T \left(\IF\right)^{-1}\vec{u}$ is maximised for any real vectors $\vec{u}$ when $M_0=M_1=\frac{1}{2}$. This shows that the best Cram\'{e}r-Rao bound for our simultaneous phase and loss estimation using a fixed number of photons is based on the SLD. Incidentally, the right logarithmic derivative bound is zero, and therefore the least informative for our problem.


\section{Optimal states for high photon numbers}
\label{sec:figs}

In Fig.6 below, we report the values of the coefficients $x_k$ obtained by numerically minimising $\Delta \theta.$ For comparison, we also present the the coefficients of the states for lossy phase estimation alone~\cite{Dorner_et_al_2009} for optimal phase estimation only. While the latter states have greater number variance, the states for simultaneous estimation, the weight $x_n$ for $n$ photons on the loss mode is higher in order to improve the estimation on $\eta.$ This comes at the cost of reduced variance, and hence reduced phase precision, as is to be expected.

\begin{widetext}

\begin{figure}[h!]
\centering
\subfigure[~$n=5$]{\includegraphics[scale=0.45]{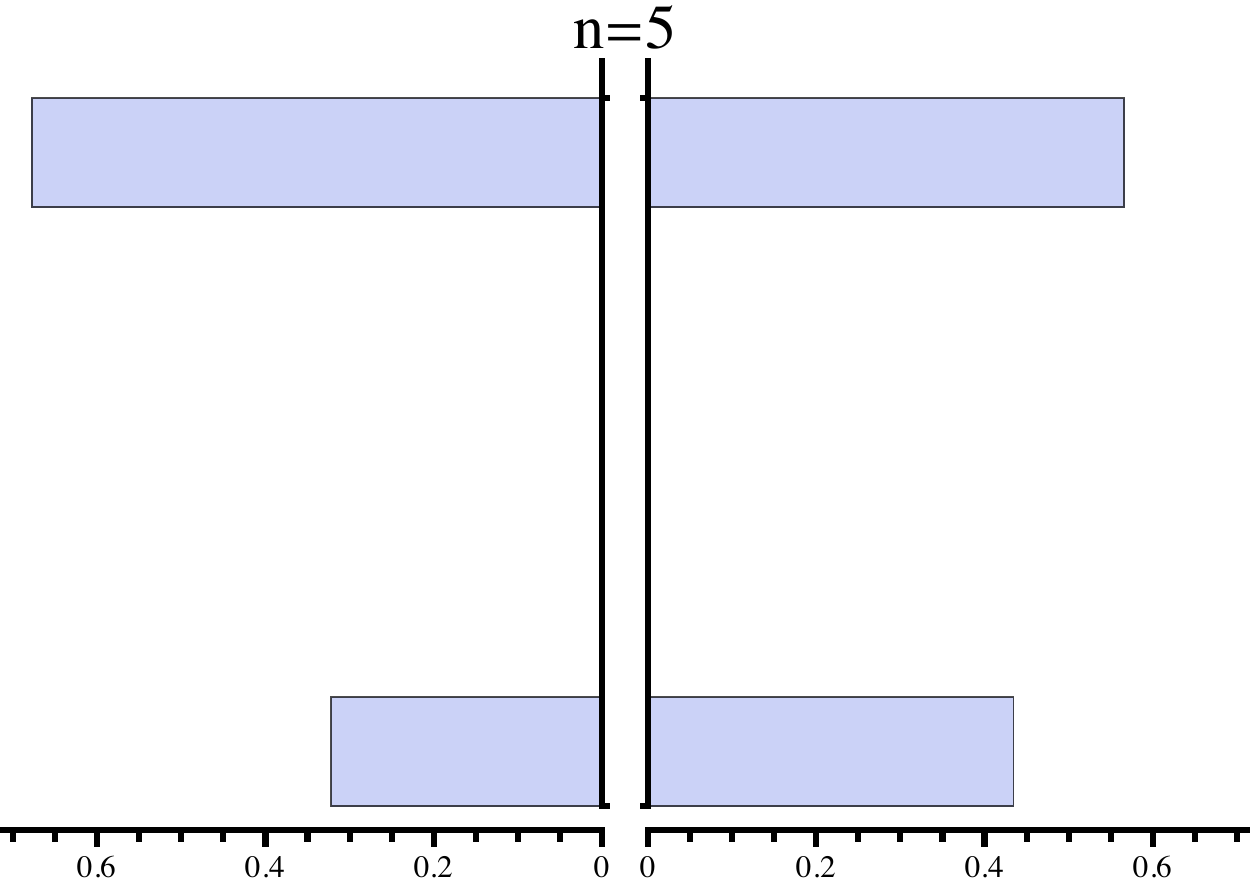}}
\subfigure[~$n=10$]{\includegraphics[scale=0.45]{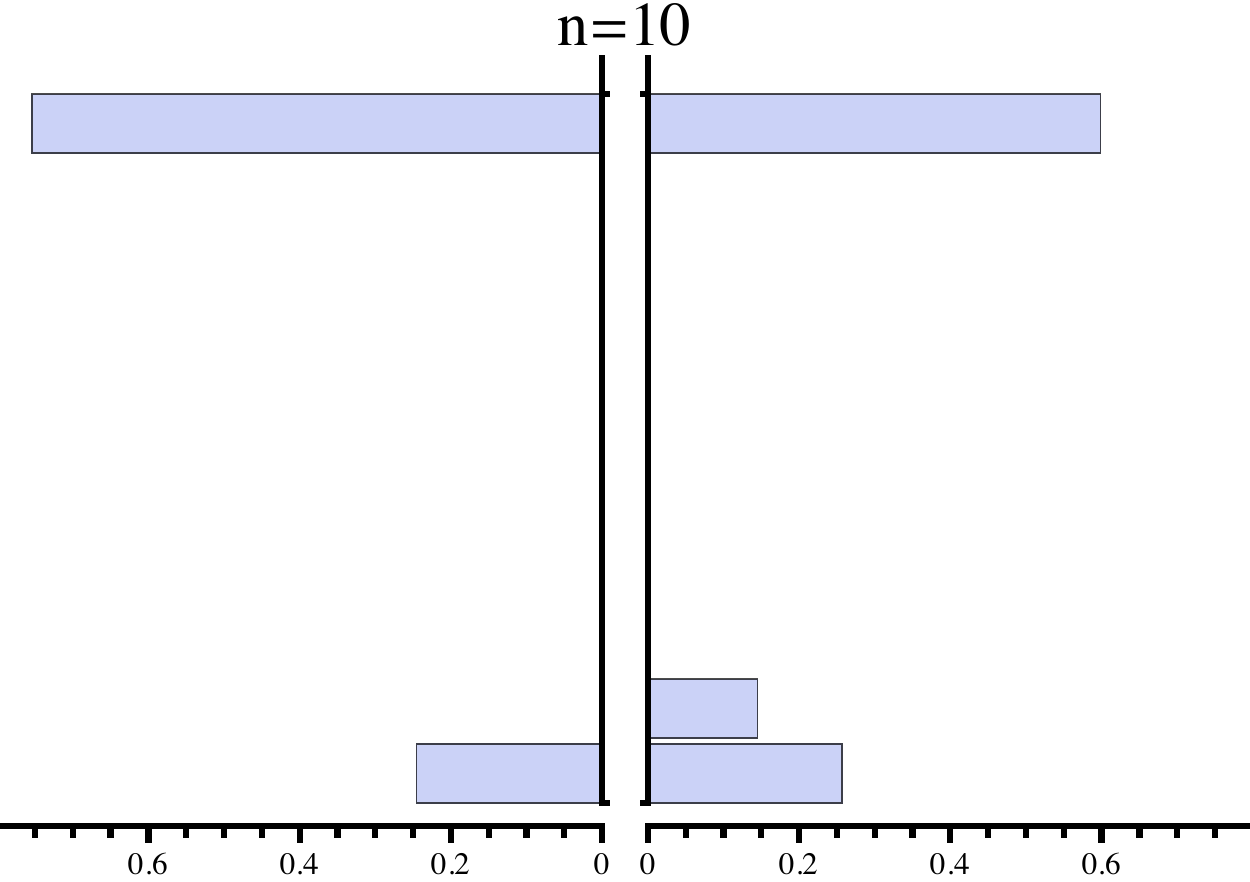}}
\subfigure[~$n=50$]{\includegraphics[scale=0.45]{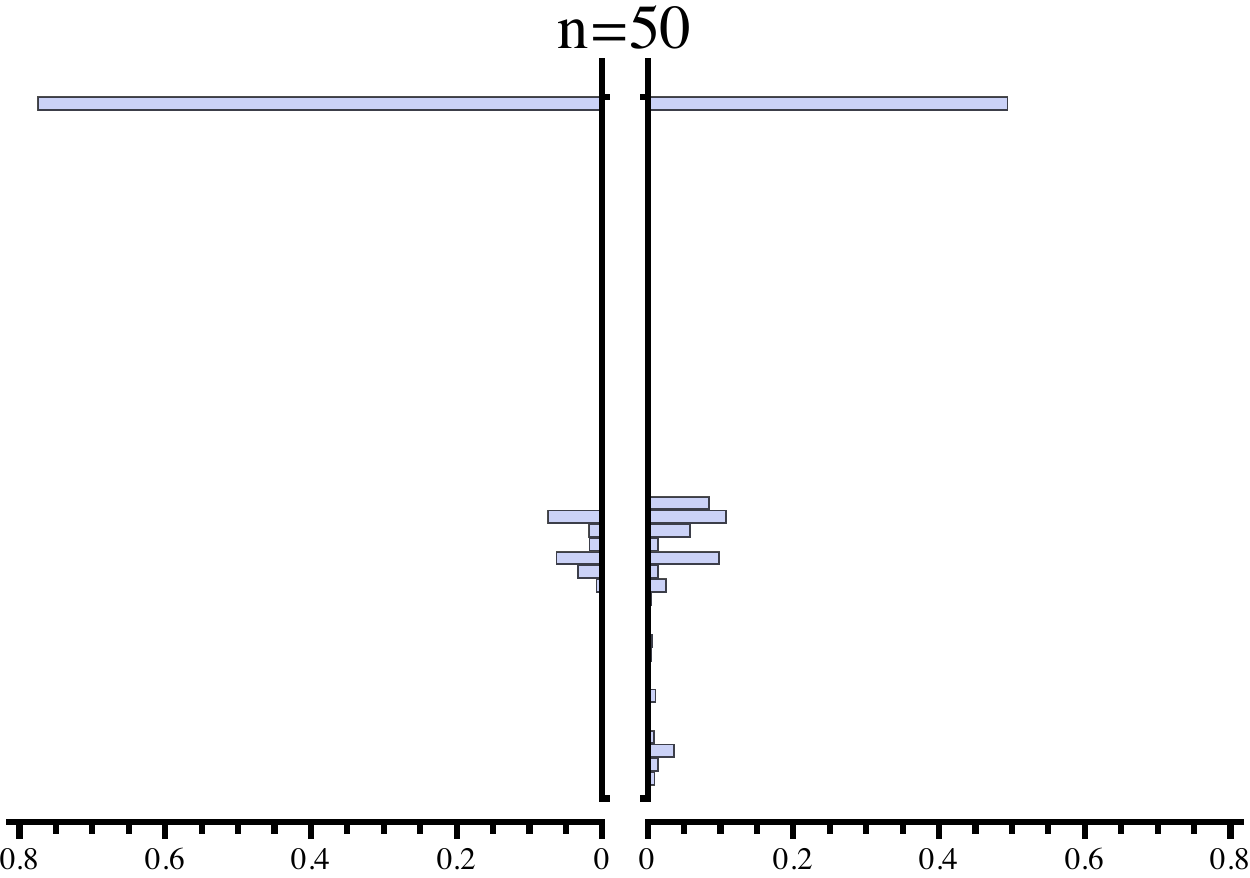}}
\subfigure[~$n=100$]{\includegraphics[scale=0.45]{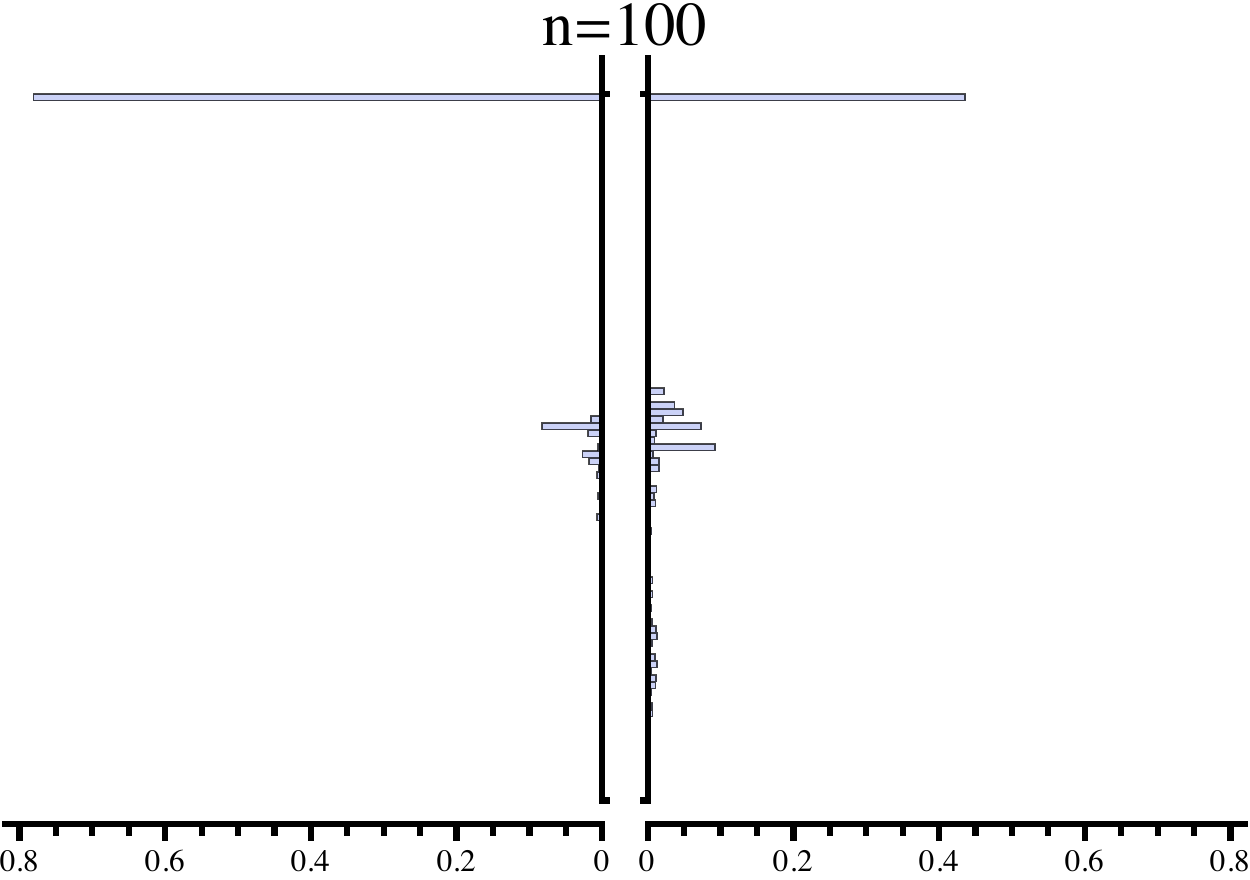}}
\subfigure[~$n=200$]{\includegraphics[scale=0.45]{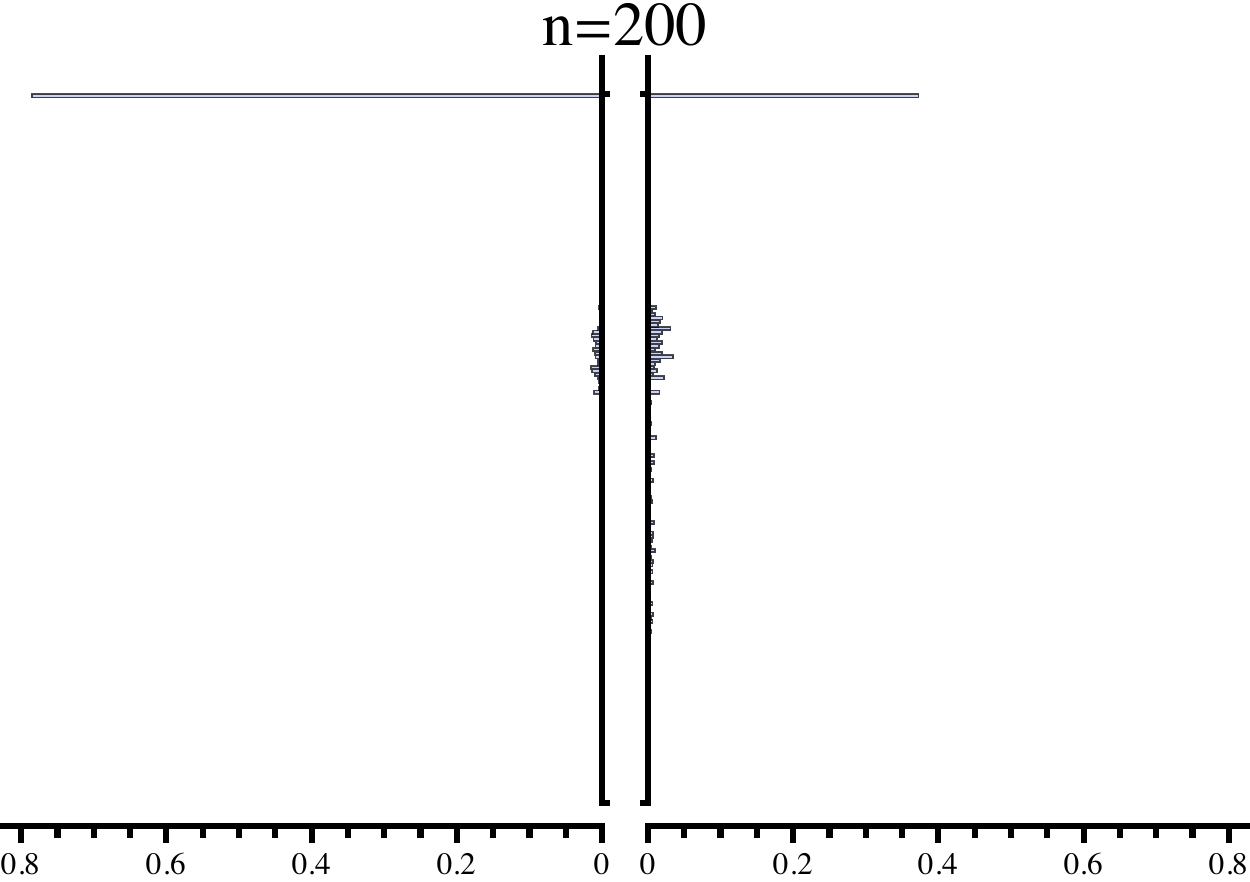}}
\caption{\label{fig:plots} The probe states with fixed number of photons $n$ for the joint estimation of phase and loss simultaneously (left), and phase only (right). The bars represent the values of $x_k=|\alpha_k|^2,$ with $k=n$ at the top to $k=0$ at the bottom. The topmost bar thus shows the contribution of all the $n$ photons in the lossy arm.}
\end{figure}

\end{widetext}

\end{document}